\documentclass[12pt]{article}
\usepackage{amssymb}
\usepackage{amsmath}
\usepackage{endnotes}
\usepackage[doublespacing]{setspace}

\input{tcilatex}
\let\footnote=\endnote

\begin{document}

\title{Identification and well-posedness in nonparametric models with
independence conditions}
\author{Victoria Zinde-Walsh\thanks{%
The support of the Social Sciences and Humanities Research Council of Canada
(SSHRC) and the \textit{Fonds\ qu\'{e}becois de la recherche sur la soci\'{e}%
t\'{e} et la culture} (FRQSC) is gratefully acknowledged. } \\
\\
McGill University and CIREQ\\
victoria.zinde-walsh@mcgill.ca}
\maketitle
\date{}

\begin{center}
\bigskip \pagebreak

{\LARGE Abstract}
\end{center}

This paper provides a nonparametric analysis for several classes of models,
with cases such as classical measurement error, regression with errors in
variables, and other models that may be represented in a form involving
convolution equations. The focus here is on conditions for existence of
solutions, nonparametric identification and well-posedness in the space $%
S^{\ast }$ of generalized functions (tempered distributions). This space
provides advantages over working in function spaces by relaxing assumptions
and extending the results to include a wider variety of models, for example
by not requiring existence of density. Classes of (generalized) functions
for which solutions exist are defined; identification conditions, partial
identification and its implications are discussed. Conditions for
well-posedness are given and the related issues of plug-in estimation and
regularization are examined.

\section{Introduction}

Many statistical and econometric models involve independence (or conditional
independence) conditions that can be expressed via convolution. Examples are
independent errors, classical measurement error and Berkson error,
regressions involving data measured with these types of errors, common
factor models and models that conditionally on some variables can be
represented in similar forms, such as a nonparametric panel data model with
errors conditionally on observables independent of the idiosyncratic
component.

Although the convolution operator is well known, this paper provides
explicitly convolution equations for a wide list of models for the first
time. In many cases the analysis in the literature takes Fourier transforms
as the starting point, e.g. characteristic functions for distributions of
random vectors (as in the famous Kotlyarski lemma, 1967). The emphasis here
on convolution equations for the models provides the opportunity to
explicitly state nonparametric classes of functions defined by the model for
which such equations hold, in particular, for densities, conditional
densities and regression functions. The statistical model may give rise to
different systems of convolution equations and may be over-identified in
terms of convolution equations; some choices may be better suited to
different situations, for example, here in Section 2 two sets of convolution
equations (4 and 4a in Table 1) are provided for the same classical
measurement error model with two measurements; it turns out that one of
those allows to relax some independence conditions, while the other makes it
possible to relax a support assumption in identification. Many of the
convolution equations derived here are based on density-weighted conditional
averages of the observables.

The main distinguishing feature is that here all the functions defined by
the model are considered within the space of generalized functions $S^{\ast
},$ the space of so-called tempered distributions (they will be referred to
as generalized functions). This is the dual space, the space of linear
continuous functionals, on the space $S$ of well-behaved functions: the
functions in $S$ are infinitely differentiable and all the derivatives go to
zero at infinity faster than any power. An important advantage of assuming
the functions are in the space of generalized functions is that in that
space any distribution function has a density (generalized function) that
continuously depends on the distribution function, so that distributions
with mass points and fractal measures have well-defined generalized
densities.

Any regular function majorized by a polynomial belongs to $S^{\ast }$; this
includes polynomially growing regression functions and binary choice
regression as well as many conditional density functions. Another advantage
is that Fourier transform is an isomorphism of this space, and thus the
usual approaches in the literature that employ characteristic functions are
also included. Details about the space $S^{\ast }$ are in Schwartz (1966)
and are summarized in Zinde-Walsh (2012).

The model classes examined here lead to convolution equations that are
similar to each other in form; the main focus of this paper is on existence,
identification, partial identification and well-posedness conditions.
Existence and uniqueness of solutions to some systems of convolution
equations in the space $S^{\ast }$ were established in Zinde-Walsh (2012).
Those results are used here to state identification in each of the models.
Identification requires examining support of the functions and generalized
functions that enter into the models; if support excludes an open set then
identification at least for some unknown functions in the model fails,
however, some isolated points or lower-dimensional manifolds where the e.g.
the characteristic function takes zero values (an example is the uniform
distribution) does not preclude identification in some of the models. This
point was made in e.g. Carrasco and Florens (2010), Evdokimov and White
(2011) and is expressed here in the context of operating in $S^{\ast }.$
Support restriction for the solution may imply that only partial
identification will be provided. However, even in partially identified
models some features of interest (see, e.g. Matzkin, 2007) could be
identified thus some questions could be addressed even in the absence of
full identification. A common example of incomplete identification which
nevertheless provides important information is Gaussian deconvolution of a
blurred image of a car obtained from a traffic camera; the filtered image is
still not very good, but the licence plate number is visible for forensics.

Well-posedness conditions are emphasized here. The well-known definition by
Hadamard (1923) defines well-posedness via three conditions: existence of a
solution, uniqueness of the solution and continuity in some suitable
topology. The first two are essentially identification. Since here we shall
be defining the functions in subclasses of $S^{\ast }$ we shall consider
continuity in the topology of this generalized functions space. This
topology is weaker than the topologies in functions spaces, such as the
uniform or $L_{p}$ topologies; thus differentiating the distribution
function to obtain a density is a well-posed problem in $S^{\ast },$ by
contrast, even in the class of absolutely continuous distributions with
uniform metric where identification for density in the space $L_{1}$ holds,
well-posedness however does not obtain (see discussion in Zinde-Walsh,
2011). But even though in the weaker topology of $S^{\ast }$ well-posedness
obtains more widely, for the problems considered here some additional
restrictions may be required for well-posedness.

Well-posedness is important for plug-in estimation since if the estimators
are in a class where the problem is well-posed they are consistent, and
conversely, if well-posedness does not hold consistency will fail for some
cases. Lack of well-posedness can be remedied by regularization, but the
price is often more extensive requirements on the model and slower
convergence. For example, in deconvolution (see e.g. Fan, 1991, and most
other papers cited here) spectral cut-off regularization is utilized; it
crucially depends on knowing the rate of the decay at infinity of the
density.

Often non-parametric identification is used to justify parametric or
semi-parametric estimation; the claim here is that well-posedness should be
an important part of this justification. The reason for that is that in
estimating a possibly misspecified parametric model, the misspecified
functions of the observables belong in a nonparametric neighborhood of the
true functions; if the model is non-parametrically identified, the unique
solution to the true model exists, but without well-posedness the solution
to the parametric model and to the true one may be far apart.

For deconvolution An and Hu (2012) demonstrate well-posedness in spaces of
integrable density functions when the measurement error has a mass point;
this may happen in surveys when probability of truthful reporting is
non-zero. The conditions for well-posedness here are provided in $S^{\ast }$%
; this then additionally does not exclude mass points in the distribution of
the mismeasured variable itself; there is some empirical evidence of mass
points in earnings and income. The results here show that in $S^{\ast }$
well-posedness holds more generally: as long as the error distribution is
not super-smooth.

The solutions for the systems of convolution equations can be used in
plug-in estimation. Properties of nonparametric plug-in estimators are based
on results on stochastic convergence in $S^{\ast }$ for the solutions that
are stochastic functions expressed via the estimators of the known functions
of the observables.

Section 2 of the paper enumerates the classes of models considered here.
They are divided into three groups: 1. measurement error models with
classical and Berkson errors and possibly an additional measurement, and
common factor models that transform into those models; 2. nonparametric
regression models with classical measurement and Berkson errors in
variables; 3. measurement error and regression models with conditional
independence. The corresponding convolution equations and systems of
equations are provided and discussed. Section 3 is devoted to describing the
solutions to the convolution equations of the models. The main mathematical
aspect of the different models is that they require solving equations of a
similar form. Section 4 provides a table of identified solutions and
discusses partial identification and well-posedness. Section 5 examines
plug-in estimation. A brief conclusion follows.

\section{Convolution equations in classes of models with independence or
conditional independence}

This section derives systems of convolution equations for some important
classes of models. The first class of model is measurement error models with
some independence (classical or Berkson error) and possibly a second
measurement; the second class is regression models with classical or Berkson
type error; the third is models with conditional independence. For the first
two classes the distributional assumptions for each model and the
corresponding convolution equations are summarized in tables; it is
indicated which of the functions are known and which unknown; a brief
discussion of each model and derivation of the convolution equations
follows. The last part of this section discusses convolution equations for
two specific models with conditional independence; one is a panel data model
studied by Evdokimov (2011), the other a regression model where independence
of measurement error of some regressors obtains conditionally on a covariate.

The general assumption made here is that all the functions in the
convolution equations belong to the space of generalized functions $S^{\ast
}.$

\textbf{Assumption 1. }\textit{All the functions defined by the statistical
model are in the space of generalized functions }$S^{\ast }.$

This space of generalized function includes functions from most of the
function classes that are usually considered, but allows for some useful
generalizations. The next subsection provides the necessary definitions and
some of the implications of working in the space $S^{\ast }.$

\subsection{The space of generalized functions $S^{\ast }.$}

The space $S^{\ast }$ is the dual space, i.e. the space of continuous linear
functionals on the space $S$ of functions. The theory of generalized
functions is in Schwartz (1966); relevant details are summarized in
Zinde-Walsh (2012). In this subsection the main definitions and properties
are reproduced.

Recall the definition of $S.$

For any vector of non-negative integers $m=(m_{1},...m_{d})$ and vector $%
t\in R^{d}$ denote by $t^{m}$ the product $t_{1}^{m_{1}}...t_{d}^{m_{d}}$
and by $\partial ^{m}$ the differentiation operator $\frac{\partial ^{m_{1}}%
}{\partial x_{1}^{m_{1}}}...\frac{\partial ^{m_{d}}}{\partial x_{d}^{m_{d}}}%
; $ $C_{\infty }$ is the space of infinitely differentiable (real or
complex-valued) functions on $R^{d}.$ The space $S\subset C_{\infty }$ of
test functions is defined as:%
\begin{equation*}
S=\left\{ \psi \in C_{\infty }(R^{d}):|t^{l}\partial ^{k}\psi (t)|=o(1)\text{
as }t\rightarrow \infty \right\} ,
\end{equation*}%
for any $k=(k_{1},...k_{d}),l=(l_{1},...l_{d}),$ where $k=(0,...0)$
corresponds to the function itself, $t\rightarrow \infty $ coordinate-wise;
thus \ the functions in $S$ go to zero at infinity faster than any power as
do their derivatives; they are rapidly decreasing functions. A sequence in $%
S $ converges if in every bounded region each $\left\vert t^{l}\partial
^{k}\psi (t)\right\vert $ converges uniformly.

Then in the dual space $S^{\ast }$ any $b\in S^{\ast }$ represents a linear
functional on $S;$ the value of this functional for $\psi \in S$ is denoted
by $\left( b,\psi \right) .$ When $b$ is an ordinary (point-wise defined)
real-valued function, such as a density of an absolutely continuous
distribution or a regression function, the value of the functional on
real-valued $\psi $ defines it and is given by 
\begin{equation*}
\left( b,\psi \right) =\int b(x)\psi (x)dx.
\end{equation*}%
If $b$ is a characteristic function it may be complex-valued, then the value
of the functional $b$ applied to $\psi \in S$ where $S$ is the space of
complex-valued functions, is 
\begin{equation*}
\left( b,\psi \right) =\int b(x)\overline{\psi (x)}dx,
\end{equation*}%
where overbar denotes complex conjugate. The integrals are taken over the
whole space $R^{d}.$

The generalized functions in the space $S^{\ast }$ are continuously
differentiable and the differentiation operator is continuous; Fourier
transforms and their inverses are defined for all $b\in S^{\ast },$ the
operator is a (continuos) isomorphism of the space $S^{\ast }.$ However,
convolutions and products are not defined for all pairs of elements of $%
S^{\ast },$ unlike, say, the space $L_{1};$ on the other hand, in $L_{1}$
differentiation is not defined and not every distribution has a density that
is an element of $L_{1}.$

Assumption 1 places no restrictions on the distributions, since in $S^{\ast
} $ any distribution function is differentiable and the differentiation
operator is continuous. The advantage of not restricting distributions to be
absolutely continuous is that mass points need not be excluded;
distributions representing fractal measures such as the Cantor distribution
are also allowed. This means that mixtures of discrete and continuous
distributions e.g. such as those examined by An and Hu (2012) for
measurement error in survey responses, some of which may be
error-contaminated, but some may be truthful leading to a mixture with a
mass point distribution are included. Moreover, in $S^{\ast }$ the case of
mass points in the distribution of the mismeasured variable is also easily
handled; in the literature such mass points are documented for income or
work hours distributions in the presence of rigidities such as unemployment
compensation rules (e.g. Green and Riddell, 1997). Fractal distributions may
arise in some situations, e.g. Karlin's (1958) example of the equilibrium
price distribution in an oligopolistic game.

For regression functions the assumption $g\in S^{\ast }$ implies that growth
at infinity is allowed but is somewhat restricted. In particular for any
ordinary point-wise defined function $b\in S^{\ast }$ the condition%
\begin{equation}
\int ...\int \Pi _{i=1}^{d}\left( (1+t_{i}^{2}\right)
^{-1})^{m_{i}}\left\vert b(t)\right\vert dt_{1}...dt_{d}<\infty ,
\label{condition}
\end{equation}%
needs to be satisfied for some non-negative valued $m_{1},...,m_{d}.$ If a
locally integrable function $g$ is such that its growth at infinity is
majorized by a polynomial, then $b\equiv g$ satisfies this condition. While
restrictive this still widens the applicability of many currently available
approaches. For example in Berkson regression the common assumption is that
the regression function be absolutely integrable (Meister, 2009); this
excludes binary choice, linear and polynomial regression functions that
belong to $S^{\ast }$ and satisfy Assumption 1. Also, it is advantageous to
allow for functions that may not belong to any ordinary function classes,
such as sums of $\delta -$functions ("sum of peaks") or (mixture) cases with
sparse parts of support, such as isolated points; such functions are in $%
S^{\ast }.$ Distributions with mass points can arise when the response to a
survey questions may be only partially contaminated; regression "sum of
peaks" functions arise e.g. in spectroscopy and astrophysics where isolated
point supports are common.

\subsection{Measurement error and related models}

Current reviews for measurement error models are in Carrol et al, (2006),
Chen et al (2011), Meister (2009).

Here and everywhere below the variables $x,z,x^{\ast },u,u_{x}$ are assumed
to be in $R^{d};y,v$ are in $R^{1};$ all the integrals are over the
corresponding space; density of $\nu $ for any $\nu $ is denoted by $f_{v};$
independence is denoted by $\bot $; expectation of $x$ conditional on $z$ is
denoted by $E(x|z).$

\subsubsection{List of models and corresponding equations}

The table below lists various models and corresponding convolution
equations. Many of the equations are derived from density weighted
conditional expectations of the observables.

Recall that for two functions, $f$ and $g$ convolution $f\ast g$ is defined
by%
\begin{equation*}
(f\ast g)\left( x\right) =\int f(w)g(x-w)dw;
\end{equation*}%
this expression is not always defined. A similar expression (with some abuse
of notation since generalized functions are not defined pointwise) may hold
for generalized functions in $S^{\ast };$ similarly, it is not always
defined. With Assumption 1 for the models considered here we show that
convolution equations given in the Tables below hold in $S^{\ast }.$

\begin{center}
\textbf{Table 1.} Measurement error models: 1. Classical measurement error;
2. Berkson measurement error; 3. Classical measurement error with additional
observation (with zero conditional mean error); 4., 4a. Classical error with
additional observation (full independence).

\begin{tabular}{|c|c|c|c|c|}
\hline
Model & $%
\begin{array}{c}
\text{Distributional} \\ 
\text{assumptions}%
\end{array}%
$ & $%
\begin{array}{c}
\text{Convolution } \\ 
\text{equations}%
\end{array}%
$ & $%
\begin{array}{c}
\text{Known} \\ 
\text{ functions}%
\end{array}%
$ & $%
\begin{array}{c}
\text{Unknown} \\ 
\text{ functions}%
\end{array}%
$ \\ \hline
\multicolumn{1}{|l|}{$\ \ \ $1.} & \multicolumn{1}{|l|}{$\ \ \ \ \ \ \ 
\begin{array}{c}
z=x^{\ast }+u \\ 
x^{\ast }\bot u%
\end{array}%
$} & \multicolumn{1}{|l|}{$\ \ \ \ \ \ \ f_{x^{\ast }}\ast f_{u}=f_{z}$} & 
\multicolumn{1}{|l|}{$\ \ \ \ \ \ \ f_{z},f_{u}$} & \multicolumn{1}{|l|}{$\
\ \ \ \ \ \ f_{x^{\ast }}$} \\ \hline
2. & $\ 
\begin{array}{c}
z=x^{\ast }+u \\ 
z\bot u%
\end{array}%
$ & $f_{z}\ast f_{-u}=f_{x^{\ast }}$ & $f_{z},f_{u}$ & $f_{x^{\ast }}$ \\ 
\hline
\multicolumn{1}{|l|}{$\ $\ 3.} & \multicolumn{1}{|l|}{$\ \ 
\begin{array}{c}
z=x^{\ast }+u; \\ 
x=x^{\ast }+u_{x} \\ 
x^{\ast }\bot u; \\ 
E(u_{x}|x^{\ast },u)=0; \\ 
E\left\Vert z\right\Vert <\infty ;E\left\Vert u\right\Vert <\infty .%
\end{array}%
$} & \multicolumn{1}{|l|}{$%
\begin{array}{c}
f_{x^{\ast }}\ast f_{u}=f_{z}; \\ 
h_{k}\ast f_{u}=w_{k}, \\ 
\text{with }h_{k}(x)\equiv x_{k}f_{x^{\ast }}(x); \\ 
k=1,2...d%
\end{array}%
$} & \multicolumn{1}{|l|}{$%
\begin{array}{c}
f_{z},w_{k}, \\ 
k=1,2...d%
\end{array}%
$} & \multicolumn{1}{|l|}{$f_{x^{\ast }}$; $f_{u}$} \\ \hline
4. & $%
\begin{array}{c}
z=x^{\ast }+u; \\ 
x=x^{\ast }+u_{x};x^{\ast }\bot u; \\ 
x^{\ast }\bot u_{x};E(u_{x})=0; \\ 
u\bot u_{x}; \\ 
E\left\Vert z\right\Vert <\infty ;E\left\Vert u\right\Vert <\infty .%
\end{array}%
$ & $%
\begin{array}{c}
f_{x^{\ast }}\ast f_{u}=f_{z}; \\ 
h_{k}\ast f_{u}=w_{k}; \\ 
f_{x^{\ast }}\ast f_{u_{x}}=f_{x}; \\ 
\text{with }h_{k}(x)\equiv x_{k}f_{x^{\ast }}(x); \\ 
k=1,2...d%
\end{array}%
$ & $%
\begin{array}{c}
f_{z}\text{, }f_{x};w;w_{k} \\ 
k=1,2...d%
\end{array}%
$ & $f_{x^{\ast }};f_{u},$ $f_{u_{x}}$ \\ \hline
4a. & $%
\begin{array}{c}
\text{Same model as 4.,} \\ 
\text{alternative} \\ 
\text{equations:}%
\end{array}%
$ & $%
\begin{array}{c}
f_{x^{\ast }}\ast f_{u}=f_{z}; \\ 
f_{u_{x}}\ast f_{-u}=w; \\ 
h_{k}\ast f_{-u}=w_{k}, \\ 
\text{with }h_{k}(x)\equiv x_{k}f_{u_{x}}(x); \\ 
k=1,2...d%
\end{array}%
$ & --"-- & --"-- \\ \hline
\end{tabular}
\end{center}

Notation: $k=1,2,...,d;$ in 3. and 4, $w_{k}=E(x_{k}f_{z}(z)|z);$ in 4a $%
w=f_{z-x};w_{k}=E(x_{k}w(z-x)|\left( z-x\right) ).$

\textbf{Theorem 1.} \textit{Under Assumption 1 for each of the models 1-4
the corresponding convolution equations of Table 1 hold in the generalized
functions space }$S^{\ast }$\textit{.}

The proof is in the derivations of the following subsection.

Assumption 1 requires considering all the functions defined by the model as
elements of the space $S^{\ast },$ but if the functions (e.g. densities, the
conditional moments) exist as regular functions, the convolutions are just
the usual convolutions of functions, on the other hand, the assumption
allows to consider convolutions for cases where distributions are not
absolutely continuous.

\subsubsection{\protect\bigskip Measurement error models and derivation of
the corresponding equations.}

1. The classical measurement error model.

The case of the classical measurement error is well known in the literature.
The concept of error independent of the variable of interest is applicable
to many problems in seismology, image processing, where it may be assumed
that the source of the error is unrelated to the signal. In e.g. Cunha et
al. (2010) it is assumed that some constructed measurement of ability of a
child derived from test scores fits into this framework. As is well-known in
regression a measurement error in the regressor can result in a biased
estimator (attenuation bias).

Typically the convolution equation%
\begin{equation*}
f_{x^{\ast }}\ast f_{u}=f_{z}
\end{equation*}%
is written for density functions when the distribution function is
absolutely continuous. The usual approach to possible non-existence of
density avoids considering the convolution and focuses on the characteristic
functions. Since density always exists as a generalized function and
convolution for such generalized functions is always defined it is possible
to write convolution equations in $S^{\ast }$ for any distributions in model
1. The error distribution (and thus generalized density $f_{u})$ is assumed
known thus the solution can be obtained by "deconvolution" (Carrol et al
(2006), Meister (2009), the review of Chen et al (2011) and papers by Fan
(1991), Carrasco and Florens(2010) among others).

2. The Berkson error model.{}

For Berkson error the convolution equation is also well-known. Berkson error
of measurement arises when the measurement is somehow controlled and the
error is caused by independent factors, e.g. amount of fertilizer applied is
given but the absorption into soil is partially determined by factors
independent of that, or students' grade distribution in a course is given in
advance, or distribution of categories for evaluation of grant proposals is
determined by the granting agency. The properties of Berkson error are very
different from that of classical error of measurement, e.g. it does not lead
to attenuation bias in regression; also in the convolution equation the
unknown function is directly expressed via the known ones when the
distribution of Berkson error is known. For discussion see Carrol et al
(2006), Meister (2009), and Wang (2004).

Models 3. and 4. The classical measurement error with another observation.

In 3., 4. in the classical measurement error model the error distribution is
not known but another observation for the mis-measured variable is
available; this case has been treated in the literature and is reviewed in
Carrol et al (2006), Chen et al \ (2011). In econometrics such models were
examined by Li and Vuong (1998), Li (2002), Schennach (2004) and
subsequently others (see e.g. the review by Chen et al, 2011). In case 3 the
additional observation contains an error that is not necessarily
independent, just has conditional mean zero.

Note that here the multivariate case is treated where arbitrary dependence
for the components of vectors is allowed. For example, it may be of interest
to consider the vector of not necessarily independent latent abilities or
skills as measured by different sections of an IQ test, or the GRE scores.

Extra measurements provide additional equations. Consider for any $k=1,...d$
the function of observables $w_{k}$ defined by density weighted expectation $%
E(x_{k}f_{z}(z)|z)$ as a generalized function; it is then determined by the
values of the functional $\left( w_{k},\psi \right) $ for every $\psi \in S.$
Note that by assumption $E(x_{k}f_{z}(z)|z)=E(x_{k}^{\ast }f_{z}(z)|z);$
then for any $\psi \in S$ the value of the functional:

\begin{eqnarray*}
(E(x_{k}^{\ast }f_{z}(z)|z),\psi ) &=&\int [\int x_{k}^{\ast }f_{x^{\ast
},z}(x^{\ast },z)dx^{\ast }]\psi (z)dz= \\
\int \int x_{k}^{\ast }f_{x^{\ast },z}(x^{\ast },z)\psi (z)dx^{\ast }dz
&=&\int \int x_{k}^{\ast }\psi (x^{\ast }+u)f_{x^{\ast },u}(x^{\ast
},u)dx^{\ast }du= \\
\int \int x_{k}^{\ast }f_{x^{\ast }}(x^{\ast })f_{u}(u)\psi (x^{\ast
}+u)dx^{\ast }du &=&(h_{k}\ast f_{u},\psi ).
\end{eqnarray*}

\bigskip The third expression is a double integral which always exists if $%
E\left\Vert x^{\ast }\right\Vert <\infty $; this is a consequence of
boundedness of the expectations of $z$ and $u.$ The fourth is a result of
change of variables $\left( x^{\ast },z\right) $ into $\left( x^{\ast
},u\right) ,$ the fifth uses independence of $x^{\ast }$and $u,$ and the
sixth expression follows from the corresponding expression for the
convolution of generalized functions (Schwartz, 1967, p.246). The conditions
of model 3 are not sufficient to identify the distribution of $u_{x};$ this
is treated as a nuisance part in model 3.

The model in 4 with all the errors and mis-measured variable independent of
each other was investigated by Kotlyarski (1967) who worked with the joint
characteristic function. In 4 consider in addition to the equations written
for model 3 another that uses the independence between $x^{\ast }$ and $%
u_{x} $ and involves $f_{u_{x}}.$

In representation 4a the convolution equations involving the density $%
f_{u_{x}}$ are obtained by applying the derivations that were used here for
the model in 3.: 
\begin{equation*}
\begin{array}{c}
z=x^{\ast }+u; \\ 
x=x^{\ast }+u_{x},%
\end{array}%
\end{equation*}%
to the model in 4 with $x-z$ playing the role of $z,$ $u_{x}$ playing the
role of $x^{\ast },$ $-u$ playing the role of $u,$ and $x^{\ast }$ playing
the role of $u_{x}.$ The additional convolution equations arising from the
extra independence conditions provide extra equations and involve the
unknown density $f_{u_{x}}.$ This representation leads to a generalization
of Kotlyarski's identification result similar to that obtained by Evdokimov
(2011) who used the joint characteristic function. The equations in 4a make
it possible to identify $f_{u},f_{u_{x}}$ ahead of $f_{x^{\ast }};$ for
identification this will require less restrictive conditions on the support
of the characteristic function for $x^{\ast }.$

\subsubsection{Some extensions}

\textbf{A. Common factor models.}

Consider a model $\tilde{z}=AU,$ with $A$ a matrix of known constants and $%
\tilde{z}$ a $m\times 1$ vector of observables, $\ U$ a vector of
unobservable variables. Usually, $A$ is a block matrix and $AU$ can be
represented via a combination of mutually independent vectors. Then without
loss of generality consider the model%
\begin{equation}
\tilde{z}=\tilde{A}x^{\ast }+\tilde{u},  \label{factormod}
\end{equation}%
where $\tilde{A}$ is a $m\times d$ known matrix of constants, $\tilde{z}$ is
a $m\times 1$ vector of observables, unobserved $x^{\ast }$ is $d\times 1$
and unobserved $\tilde{u}$ is $m\times 1.$ If the model $\left( \ref%
{factormod}\right) $ can be transformed to model 3 considered above, then $%
x^{\ast }$ will be identified whenever identification holds for model 3.
Once some components are identified identification of other factors could be
considered sequentially.

\textbf{Lemma 1. }\textit{If in }$\left( \ref{factormod}\right) $ \textit{%
the vectors }$x^{\ast }$\textit{\ and }$\tilde{u}$\textit{\ are independent
and all the components of the vector }$\tilde{u}$\textit{\ are mean
independent of each other and are mean zero and the matrix }$A$ \textit{can
be partitioned after possibly some permutation of rows as }$\left( 
\begin{array}{c}
A_{1} \\ 
A_{2}%
\end{array}%
\right) $\textit{\ with }$rankA_{1}=rankA_{2}=d,$\textit{\ then the model }$%
\left( \ref{factormod}\right) $\textit{\ implies model 3.}

Proof. Define $z=T_{1}\tilde{z},$ where conformably to the partition of $A$
the partitioned $T_{1}=\left( 
\begin{array}{c}
\tilde{T}_{1} \\ 
0%
\end{array}%
\right) ,$ with $\tilde{T}_{1}A_{1}x^{\ast }=x^{\ast }$ (such a $\tilde{T}%
_{1}$ always exists by the rank condition); then $z=x^{\ast }+u,$ where $%
u=T_{1}\tilde{u}$ is independent of $x^{\ast }.$ Next define $T_{2}=\left( 
\begin{array}{c}
0 \\ 
\tilde{T}_{2}%
\end{array}%
\right) $ similarly with $\tilde{T}_{2}A_{2}x^{\ast }=x^{\ast }$.

Then $x=T_{2}\tilde{z}$ is such that $x=x^{\ast }+u_{x},$ where $u_{x}=T_{2}%
\tilde{u}$ and does not include any components from $u.$ This implies $%
Eu_{x}|(x^{\ast },u)=0.$ Model 3 holds. $\blacksquare $

Here dependence in components of $x^{\ast }$ is arbitrary. A general
structure with subvectors of $U$ independent of each other but with
components which may be only mean independent (as $\tilde{u}$ here) or
arbitrarily dependent (as in $x^{\ast })$ is examined by Ben-Moshe (2012).
Models of linear systems with full independence were examined by e.g. Li and
Vuong (1998). These models lead to systems of first-order differential
equations for the characteristic functions.

It may be that there are no independent components $x^{\ast }$ and $\tilde{u}
$ for which the conditions of Lemma 1 are satisfied. Bonhomme and Robin
(2010) proposed to consider products of the observables to increase the
number of equations in the system and analyzed conditions for
identification; Ben-Moshe (2012) provided necessary and sufficient
conditions under which this strategy leads to identification when there may
be some dependence.

\textbf{B. Error correlations with more observables.}

The extension to non-zero $E(u_{x}|z)$ in model 3 is trivial if this
expectation is a known function. A more interesting case results if the
errors $u_{x}$ and $u$ are related, e.g. 
\begin{equation*}
u_{x}=\rho u+\eta ;\eta \bot z.
\end{equation*}

With an unknown parameter (or function of observables) $\rho $ if more
observations are available more convolution equations can be written to
identify all the unknown functions. Suppose that additionally a observation $%
y$ is available with 
\begin{eqnarray*}
y &=&x^{\ast }+u_{y}; \\
u_{y} &=&\rho u_{x}+\eta _{1};\eta _{1}\bot ,\eta ,z.
\end{eqnarray*}%
Without loss of generality consider the univariate case and define $%
w_{x}=E(xf(z)|z);w_{y}=E(yf(z)|z).\,\ $Then the system of convolution
equations expands to

\begin{equation}
\left\{ 
\begin{array}{ccc}
f_{x^{\ast }}\ast f_{u} &  & =w; \\ 
(1-\rho )h_{x^{\ast }}\ast f_{u} & +\rho zf(z) & =w_{x}; \\ 
(1-\rho ^{2})h_{x^{\ast }}\ast f_{u} & +\rho ^{2}zf(z) & =w_{y}.%
\end{array}%
\right.  \label{ar(1)}
\end{equation}

The three equations have three unknown functions, $f_{x^{\ast }},f_{u}$ and $%
\rho .$ Assuming that support of $\rho $ does not include the point 1, $\rho 
$ can be expressed as a solution to a linear algebraic equation derived from
the two equations in $\left( \ref{ar(1)}\right) $ that include $\rho :$ 
\begin{equation*}
\rho =(w_{x}-zf(z))^{-1}\left( w_{y}-w_{x}\right) .
\end{equation*}

\subsection{Regression models with classical and Berkson errors and the
convolution equations}

\subsubsection{The list of models}

The table below provides several regression models and the corresponding
convolution equations involving density weighted conditional expectations.

\begin{center}
Table 2. Regression models: 5. Regression with classical measurement error
and an additional observation; 6. Regression with Berkson error ($x,y,z$ are
observable); 7. Regression with zero mean measurement error and Berkson
instruments.
\end{center}

\begin{tabular}{|c|c|c|c|c|}
\hline
Model & $%
\begin{array}{c}
\text{Distributional} \\ 
\text{assumptions}%
\end{array}%
$ & $%
\begin{array}{c}
\text{Convolution } \\ 
\text{equations}%
\end{array}%
$ & $%
\begin{array}{c}
\text{Known} \\ 
\text{ functions}%
\end{array}%
$ & $%
\begin{array}{c}
\text{Unknown} \\ 
\text{ functions}%
\end{array}%
$ \\ \hline
\multicolumn{1}{|l|}{$\ \ $5.} & \multicolumn{1}{|l|}{$\ \ 
\begin{array}{c}
y=g(x^{\ast })+v \\ 
z=x^{\ast }+u; \\ 
x=x^{\ast }+u_{x} \\ 
x^{\ast }\bot u;E(u)=0; \\ 
E(u_{x}|x^{\ast },u)=0; \\ 
E(v|x^{\ast },u,u_{x})=0.%
\end{array}%
$} & \multicolumn{1}{|l|}{$%
\begin{array}{c}
f_{x^{\ast }}\ast f_{u}=f_{z}; \\ 
\left( gf_{x^{\ast }}\right) \ast f_{u}=w, \\ 
h_{k}\ast f_{u}=w_{k}; \\ 
\text{with }h_{k}(x)\equiv x_{k}g(x)f_{x^{\ast }}(x); \\ 
k=1,2...d%
\end{array}%
$} & \multicolumn{1}{|l|}{$f_{z};$ $w;w_{k}$} & \multicolumn{1}{|l|}{$%
f_{x^{\ast }}$; $f_{u}$; $g.$} \\ \hline
\multicolumn{1}{|l|}{$\ \ $6.} & \multicolumn{1}{|l|}{$%
\begin{array}{c}
y=g(x)+v \\ 
z=x+u;E(v|z)=0; \\ 
z\bot u;E(u)=0.%
\end{array}%
$} & \multicolumn{1}{|l|}{$\ \ \ \ \ 
\begin{array}{c}
f_{x}=f_{-u}\ast f_{z}; \\ 
g\ast f_{-u}=w%
\end{array}%
$} & \multicolumn{1}{|l|}{$\ f_{z};f_{x},w$} & \multicolumn{1}{|l|}{$f_{u}$; 
$g.$} \\ \hline
\multicolumn{1}{|l|}{$\ \ $7.} & \multicolumn{1}{|l|}{$\ \ 
\begin{array}{c}
y=g(x^{\ast })+v; \\ 
x=x^{\ast }+u_{x}; \\ 
z=x^{\ast }+u;z\bot u; \\ 
E(v|z,u,u_{x})=0; \\ 
E(u_{x}|z,v)=0.%
\end{array}%
$} & \multicolumn{1}{|l|}{$%
\begin{array}{c}
g\ast f_{u}=w; \\ 
h_{k}\ast f_{u}=w_{k}, \\ 
\text{with }h_{k}(x)\equiv x_{k}g(x); \\ 
k=1,2...d%
\end{array}%
$} & \multicolumn{1}{|l|}{$w,w_{k}$} & \multicolumn{1}{|l|}{$f_{u}$; $g.$}
\\ \hline
\end{tabular}

\bigskip Notes. Notation: $k=1,2...d;$ in model 5.$%
w=E(yf_{z}(z)|z);w_{k}=E(x_{k}f_{z}(z)|z);$ in model 6. $w=E(y|z);$ in model
7. $w=E(y|z);w_{k}=E(x_{k}y|z).$

\textbf{Theorem 2.} \textit{Under Assumption 1 for each of the models 5-7
the corresponding convolution equations hold.}

\bigskip The proof is in the derivations of the next subsection.

\subsubsection{\protect\bigskip Discussion of the regression models and
derivation of the convolution equations.}

5. The nonparametric regression model with classical measurement error and
an additional observation.

This type of model was examined by Li (2002) and Li and Hsiao (2004); the
convolution equations derived here provide a convenient representation.
Often models of this type were considered in semiparametric settings.
Butucea and Taupin (2008) (extending the earlier approach by Taupin, 2001)
consider a regression function known up to a finite dimensional parameter
with the mismeasured variable observed with independent error where the
error distribution is known. Under the latter condition the model 5 here
would reduce to the two first equations 
\begin{equation*}
f_{x^{\ast }}\ast f_{u}=f_{z};\text{ }\left( gf_{x^{\ast }}\right) \ast
f_{u}=w,
\end{equation*}%
where $f_{u}$ is known and two unknown functions are $g$ (here
nonparametric) and $f_{x^{\ast }}.$

The model 5 incorporates model 3 for the regressor and thus the convolution
equations from that model apply. An additional convolution equation is
derived here; it is obtained from considering the value of the density
weighted conditional expectation in the dual space of generalized functions, 
$S^{\ast },$ applied to arbitrary $\psi \in S,$%
\begin{equation*}
(w,\psi )=(E(f(z)y|z),\psi )=(E(f(z)g(x^{\ast })|z),\psi );
\end{equation*}%
this equals 
\begin{eqnarray*}
&&\int \int g(x^{\ast })f_{x^{\ast },z}(x^{\ast },z)\psi (z)dx^{\ast }dz \\
&=&\int \int g(x^{\ast })f_{x^{\ast },u}(x^{\ast },u)\psi (x^{\ast
}+u)dx^{\ast }du \\
&=&\int g(x^{\ast })f_{x^{\ast }}(x^{\ast })f_{u}(u)dx^{\ast }\psi (x^{\ast
}+u)dx^{\ast }du=((gf_{x^{\ast }})\ast f_{u},\psi ).
\end{eqnarray*}

Conditional moments for the regression function need not be integrable or
bounded functions of $z$; we require them to be in the space of generalized
functions $S^{\ast }.$

6. Regression with Berkson error.

This model may represent the situation when the regressor (observed) $x$ is
correlated with the error $v,$ but $z$ is a (vector) possibly representing
an instrument uncorrelated with the regression error.

Then as is known in addition to the Berkson error convolution equation the
equation 
\begin{equation*}
w=E(y|z)=E(g(x)|z)=\int g(x)\frac{f_{x,z}(x,z)}{f_{z}(z)}dx=\int
g(z-u)f_{u}(u)dx=g\ast f_{u}
\end{equation*}%
holds. This is stated in Meister (2008); however, the approach there is to
consider $g$ to be absolutely integrable so that convolution can be defined
in the $L_{1}$ space. Here by working in the space of generalized functions $%
S^{\ast }$ a much wider nonparametric class of functions that includes
regression functions with polynomial growth is allowed.

7. Nonparametric regression with error in the regressor, where Berkson type
instruments are assumed available.

This model was proposed by Newey (2001), examined in the univarite case by
Schennach (2007) and Zinde-Walsh (2009), in the multivariate case in
Zinde-Walsh (2012), where the convolution equations given here in Table 2
were derived.

\subsection{\textbf{Convolution equations in models with conditional
independence conditions.}}

All the models 1-7 can be extended to include some additional variables
where conditionally on those variables, the functions in the model (e.g.
conditional distributions) are defined and all the model assumptions hold
conditionally.

Evdokimov (2011) derived the conditional version of the model 4 from a very
general nonparametric panel data model. Model 8 below describes the panel
data set-up and how it transforms to conditional model 4 and 4a and possibly
model 3 with relaxed independence condition (if the focus is on identifying
the regression function).

Model 8. Panel data model with conditional independence.

Consider a two-period panel data model with an unknown regression function $%
m $ and an idiosyncratic (unobserved) $\alpha :$ 
\begin{eqnarray*}
Y_{i1} &=&m(X_{i1},\alpha _{i})+U_{i1}; \\
Y_{i2} &=&m(X_{i2},\alpha _{i})+U_{i2}.
\end{eqnarray*}

To be able to work with various conditional characteristic functions
corresponding assumptions ensuring existence of the conditional
distributions need to be made and in what follows we assume that all the
conditional density functions and moments exist as generalized functions in $%
S^{\ast }$.

In Evdokimov (2011) independence (conditional on the corresponding period $%
X^{\prime }s)$ of the regression error from $\alpha ,$ and from the $%
X^{\prime }s$ and error of the other period is assumed: 
\begin{equation*}
f_{t}=f_{Uit}|_{X_{it},\alpha
_{i},X_{i(-t)},U_{i(-t)}}(u_{t}|x,...)=f_{Uit}|_{X_{it}}(u_{t}|x),t=1,2
\end{equation*}%
with $f_{\cdot |\cdot }$ denoting corresponding conditional densities.
Conditionally on $X_{i2}=X_{i1}=x$ the model takes the form 4%
\begin{equation*}
\begin{array}{c}
z=x^{\ast }+u; \\ 
x=x^{\ast }+u_{x}%
\end{array}%
\end{equation*}%
with $z$ representing $Y_{1},x$ representing $Y_{2},$ $x^{\ast }$ standing
in for $m(x,\alpha ),$ $u$ for $U_{1}$ and $u_{x}$ for $U_{2}.$ The
convolution equations derived here for 4 or 4a now apply to conditional
densities.

The convolution equations in 4a are similar to Evdokimov; they allow for
equations for $f_{u},$ $f_{u_{x}}$ that do not rely on $f_{x^{\ast }}.$ The
advantage of those lies in the possibility of identifying the conditional
error distributions without placing the usual non-zero restrictions on the
characteristic function of $x^{\ast }$ (that represents the function $m$ for
the panel model).

The panel model can be considered with relaxed independence assumptions.
Here in the two-period model we look at forms of dependence that assume zero
conditional mean of the second period error, rather than full independence
of the first period error: 
\begin{eqnarray*}
f_{Ui1}|_{X_{i1},\alpha _{i},X_{i2},Ui2}(u_{t}|x,...)
&=&f_{Ui1}|_{Xi1}(u_{t}|x); \\
E(U_{i2}|X_{i1},\alpha _{i},X_{i2},U_{i1}) &=&0; \\
f_{Ui2}|_{\alpha _{i},X_{i2}=X_{i1}=x}(u_{t}|x,...)
&=&f_{Ui2}|_{Xi2}(u_{t}|x).
\end{eqnarray*}%
Then the model maps into the model 3 with the functions in the convolution
equations representing conditional densities and allows to identify
distribution of $x^{\ast }$ (function $\ m$ in the model). But the
conditional distribution of the second-period error in this set-up is not
identified.

Evdokimov introduced parametric AR(1) or MA(1) dependence in the errors $U$
and to accommodate that extended the model to three periods. Here this would
lead in the AR case to the equations in $\left( \ref{ar(1)}\right) .$

Model 9. Errors in variables regression with classical measurement error
conditionally on covariates.

Consider the regression model 
\begin{equation*}
y=g(x^{\ast },t)+v,
\end{equation*}%
with a measurement of unobserved $x^{\ast }$ given by $\ \tilde{z}=x^{\ast }+%
\tilde{u},$ with $x^{\ast }\bot \tilde{u}$ conditionally on $t$. Assume that 
$E(\tilde{u}|t)=0$ and that $E(v|x^{\ast },t)=0.$ Then redefining all the
densities and conditional expectations to be conditional on $t$ we get the
same system of convolution equations as in Table 2 for model 5 with the
unknown functions now being conditional densities and the regression
function, $g.$

Conditioning requires assumptions that provide for existence of conditional
distribution functions in $S^{\ast }$.

\section{\textbf{Solutions for the models.}}

\subsection{Existence of solutions}

To state results for nonparametric models it is important first to clearly
indicate the classes of functions where the solution is sought. Assumption 1
requires that all the (generalized) functions considered are elements in the
space of generalized functions $S^{\ast }.$ This implies that in the
equations the operation of convolution applied to the two functions from $%
S^{\ast }$ provides an element in the space $S^{\ast }.$ This subsection
gives high level assumptions on the nonparametric classes of the unknown
functions where the solutions can be sought: any functions from these
classes that enter into the convolution provide a result in $S^{\ast }.$

No assumptions are needed for existence of convolution and full generality
of identification conditions in models 1,2 where the model assumptions imply
that the functions represent generalized densities. For the other models
including regression models convolution is not always defined in $S^{\ast }.$
Zinde-Walsh (2012) defines the concept of convolution pairs of classes of
functions in $S^{\ast }$ where convolution can be applied.

To solve the convolution equations a Fourier transform is usually employed,
so that e.g. one transforms generalized density functions into
characteristic functions. Fourier transform is an isomorphism of the space $%
S^{\ast }.$ The Fourier transform of a generalized function $a\in S^{\ast }$%
, $Ft(a),$ is defined as follows. For any $\psi \in S,$ as usual $Ft(\psi
)(s)=\int \psi (x)e^{isx}dx;$ then the functional $Ft(a)$ is defined by%
\begin{equation*}
(Ft(a),\psi )\equiv (a,Ft(\psi )).
\end{equation*}%
The advantage of applying Fourier transform is that integral convolution
equations transform into algebraic equations when the "exchange formula"
applies:%
\begin{equation}
a\ast b=c\Longleftrightarrow Ft(a)\cdot Ft(b)=Ft(c).  \label{exchange}
\end{equation}%
In the space of generalized functions $S^{\ast },$ the Fourier transform and
inverse Fourier transform always exist. As shown in Zinde-Walsh (2012) there
is a dichotomy between convolution pairs of subspaces in $S^{\ast }$ and the
corresponding product pairs of subspaces of their Fourier transforms.

The classical pairs of spaces (Schwartz, 1966) are the convolution pair $%
\left( S^{\ast },O_{C}^{\ast }\right) $ and the corresponding product pair $%
\left( S^{\ast },O_{M}\right) ,$ where $O_{C}^{\ast }$ is the subspace of $%
S^{\ast }$ that contains rapidly decreasing (faster than any polynomial)
generalized functions and $\mathit{O}_{M}$ is the space of infinitely
differentiable functions with every derivative growing no faster than a
polynomial at infinity. These pairs are important in that no restriction is
placed on one of the generalized functions that could be any element of
space $S^{\ast }$; the other belongs to a space that needs to be
correspondingly restricted. A disadvantage of the classical pairs is that
the restriction is fairly severe, for example, the requirement that a
characteristic function be in $O_{M}\,\ $implies existence of all moments
for the random variable. Relaxing this restriction would require placing
constraints on the other space in the pair; Zinde-Walsh (2012) introduces
some pairs that incorporate such trade-offs.

In some models the product of a function with a component of the vector of
arguments is involved,such as $d(x)=x_{k}a(x),$ then for Fourier transforms $%
Ft(d)\left( s\right) =-i\frac{\partial }{\partial s_{k}}Ft(a)(s);$ the
multiplication by a variable is transformed into ($-i)$ times the
corresponding partial derivative. Since the differentiation operators are
continuous in $S^{\ast }$ this transformation does not present a problem.

\textbf{Assumption 2.} \textit{The functions }$a\in A,b\in B,$\textit{\ are
such that }$\left( A,B\right) $\textit{\ form a convolution pair in }$%
S^{\ast }$\textit{.}

Equivalently, $Ft(a),$ $Ft(b)$ are in the corresponding product pair of
spaces.

Assumption 2 is applied to model 1 for $a=f_{x^{\ast }},b=f_{u};$ to model 2
with $a=f_{z},b=f_{u};$ to model 3 with $a=f_{x^{\ast }},b=f_{u}$ and with $%
a=h_{k},b=f_{u},$ for all $k=1,...,d;$ to model 4a for $a=f_{x^{\ast }},$ or 
$f_{u_{x}},$ or $h_{k}$ for all $k$ and $b=f_{u};$ to model 5 with $%
a=f_{x^{\ast }},$ or $gf_{x^{\ast }},$ or $h_{k}f_{x^{\ast }}$ and $b=f_{u};$
to model 6 with $a=f_{z},$ or $g$ and $b=f_{u};$ to model 7 with $a=g$ or $%
h_{k}$ and $b=f_{u}.$

Assumption 2 is a high-level assumption that is a sufficient condition for a
solution to the models 1-4 and 6-7 to exist. Some additional conditions are
needed for model 5 and are provided below.

Assumption 2 is automatically satisfied for generalized density functions,
so is not needed for models 1 and 2. Denote by $\bar{D}\subset S^{\ast }$
the subset of generalized derivatives of distribution functions
(corresponding to Borel probability measures in $R^{d}$) then in models 1
and 2 $A=B=\bar{D};$ and for the characteristic functions there are
correspondingly no restrictions; denote the set of all characteristic
functions, $Ft\left( \bar{D}\right) \subset S^{\ast },$ by $\bar{C}.$

Below a (non-exhaustive) list of nonparametric classes of generalized
functions that provide sufficient conditions for existence of solutions to
the models here is given. The classes are such that they provide minimal or
often no restrictions on one of the functions and restrict the class of the
other in order that the assumptions be satisfied.

In models 3 and 4 the functions $h_{k}$ are transformed into derivatives of
continuous characteristic functions. An assumption that either the
characteristic function of $x^{\ast }$ or the characteristic function of $u$
be continuously differentiable is sufficient, without any restrictions on
the other to ensure that Assumption 2 holds. Define the subset of all
continuously differentiable characteristic functions by $\bar{C}^{(1)}.$

In model 5 equations involve a product of the regression function $g$ with $%
f_{x^{\ast }}.$ Products of generalized functions in $S^{\ast }$ do not
always exist and so additional restrictions are needed in that model. If $g$
is an arbitrary element of $S^{\ast },$ then for the product to exist, $%
f_{x^{\ast }}$ should be in $\mathit{O}_{M}$. On the other hand, if $%
f_{x^{\ast }}$ is an arbitrary generalized density it is sufficient that $g$
and $h_{k}$ belong to the space of $d$ times continuously differentiable
functions with derivatives that are majorized by polynomial functions for $%
gf_{x^{\ast }},h_{k}f_{x^{\ast }}$ to be elements of $S^{\ast }.$ Indeed,
the value of the functional $h_{k}f_{x^{\ast }}$ for an arbitrary $\psi \in
S $ is defined by%
\begin{equation*}
(h_{k}f_{x^{\ast }},\psi )=\left( -1\right) ^{d}\int F_{x^{\ast
}}(x)\partial ^{(1,...,1)}(h_{k}(x)\psi (x))dx;
\end{equation*}%
here $F$ is the distribution (ordinary bounded) function and this integral
exists because $\psi $ and all its derivatives go to zero at infinity faster
than any polynomial function. Denote by $\bar{S}^{B,1}$ the space of
continuously differentiable functions $g\in S^{\ast }$ such that the
functions $h_{k}(x)=x_{k}g(x)$ are also continuously differentiable with all
derivatives majorized by polynomial functions$.$ Since the products are in $%
S^{\ast }$ then the Fourier transforms of the products are defined in $%
S^{\ast }.$ Further restrictions requiring the Fourier transforms of the
products $gf_{x^{\ast }}$\ and $h_{k}f_{x^{\ast }}$ to be continuously
differentiable functions in $S^{\ast }$ would remove any restrictions on $%
f_{u}$ for the convolution to exist. Denote the space of all continuously
differentiable functions in $S^{\ast }$ by $\bar{S}^{(1)}.$

If $g$ is an ordinary function that represents a regular element in $S^{\ast
}$ the infinite differentiability condition on $f_{x^{\ast }}$ can be
relaxed to simply requiring continuous first derivatives.

In models 6 and 7 if the generalized density function for the error, $f_{u},$
decreases faster than any polynomial (all moments need to exist for that),
so that $f_{u}\in \mathit{O}_{C}^{\ast },$ \ then $g$ could be any
generalized function in $S^{\ast };$ this will of course hold if $f_{u}$ has
bounded support. Generally, the more moments the error is assumed to have,
the fewer restrictions on the regression function $g$ are needed to satisfy
the convolution equations of the model and the exchange formula. The models
6, 7 satisfy the assumptions for any error $u$ when support of generalized
function $g$ is compact (as for the "sum of peaks"), then $g\in E^{\ast
}\subset S^{\ast },$ where $E^{\ast }$ is the space of generalized functions
with compact support. More generally the functions $g$ and all the $h_{k}$
could belong to the space $\mathit{O}_{C}^{\ast }$ of generalized functions
that decrease at infinity faster than any polynomial, and still no
restrictions need to be placed on $u.$

Denote for any generalized density function $f_{\cdot }$ the corresponding
characteristic function, $Ft(f_{\cdot }),$ by $\phi _{\cdot }.$ Denote
Fourier transform of the (generalized) regression function $g,$ $Ft(g),$ by $%
\gamma .$

The following table summarizes some fairly general sufficient conditions on
the models that place restrictions on the functions themselves or on the
characteristic functions of distributions in the models that will ensure
that Assumption 2 is satisfied and a solution exists. The nature of these
assumptions is to provide restrictions on some of the functions that allow
the others to be completely unrestricted for the corresponding model.

\textbf{Table 3.} Some nonparametric classes of generalized functions for
which the convolution equations of the models are defined in $S^{\ast }$.

\begin{tabular}{|c|c|c|}
\hline
Model & Sufficient & assumptions \\ \hline
1 & no restrictions: & $\phi _{x^{\ast }}\in \bar{C};\phi _{u}\in \bar{C}$
\\ \hline
2 & no restrictions: & $\phi _{x^{\ast }}\in \bar{C};\phi _{u}\in \bar{C}$
\\ \hline
& Assumptions A & Assumptions B \\ \hline
\multicolumn{1}{|l|}{$\ \ \ \ $3} & \multicolumn{1}{|l|}{any$\ \ \phi
_{x^{\ast }}\in \bar{C};\phi _{u}\in \bar{C}^{(1)}$} & \multicolumn{1}{|l|}{
any $\phi _{u}\in \bar{C};\phi _{x^{\ast }}\in \bar{C}^{(1)}$} \\ \hline
4 & any $\phi _{u_{x}},\phi _{x^{\ast }}\in \bar{C};\phi _{u}\in \bar{C}%
^{(1)}$ & any $\phi _{u},\phi _{x^{\ast }}\in \bar{C};\phi _{u_{x}}\in \bar{C%
}^{(1)}$ \\ \hline
4a & any $\phi _{u_{x}},\phi _{x^{\ast }}\in \bar{C};\phi _{u}\in \bar{C}%
^{(1)}$ & any $\phi _{u},\phi _{u_{x}}\in \bar{C};\phi _{x^{\ast }}\in \bar{C%
}^{(1)}$ \\ \hline
\multicolumn{1}{|l|}{$\ \ \ \ $5} & \multicolumn{1}{|l|}{any $g\in S^{\ast
};f_{x^{\ast }}\in O_{M};f_{u}\in O_{C}^{\ast }$} & \multicolumn{1}{|l|}{$\ $%
any $\ f_{x^{\ast }}\in \bar{D};\ g,h_{k}\in \bar{S}^{B,1};f_{u}\in
O_{C}^{\ast }$} \\ \hline
\multicolumn{1}{|l|}{$\ \ \ \ $6} & \multicolumn{1}{|l|}{any$\ g\in S^{\ast
};f_{u}\in O_{C}^{\ast }$} & \multicolumn{1}{|l|}{$\ g\in O_{C}^{\ast };$
any $f_{u}:\phi _{u}\in \bar{C}$} \\ \hline
7 & any $g\in S^{\ast };f_{u}\in O_{C}^{\ast }$ & $g\in O_{C}^{\ast };$ any $%
f_{u}:\phi _{u}\in \bar{C}$ \\ \hline
\end{tabular}

The next table states the equations and systems of equations for Fourier
transforms that follow from the convolution equations.

\textbf{Table 4.} The form of the equations for the Fourier transforms:

\begin{tabular}{|c|c|c|}
\hline
Model & Eq's for Fourier transforms & Unknown functions \\ \hline
1 & $\phi _{x^{\ast }}\phi _{u}=\phi _{z};$ & $\phi _{x^{\ast }}$ \\ \hline
2 & $\phi _{x^{\ast }}=\phi _{z}\phi _{-u};$ & $\phi _{x^{\ast }}$ \\ \hline
3 & $\left\{ 
\begin{array}{c}
\phi _{x^{\ast }}\phi _{u}=\phi _{z}; \\ 
\left( \phi _{x^{\ast }}\right) _{k}^{\prime }\phi _{u}=\varepsilon
_{k},k=1,...,d.%
\end{array}%
\right. $ & $\phi _{x^{\ast }},\phi _{u}$ \\ \hline
4 & $\left\{ 
\begin{array}{c}
\phi _{x^{\ast }}\phi _{u}=\phi _{z}; \\ 
\left( \phi _{x^{\ast }}\right) _{k}^{\prime }\phi _{u}=\varepsilon
_{k},k=1,...,d; \\ 
\phi _{x^{\ast }}\phi _{u_{x}}=\phi _{x}.%
\end{array}%
\right. $ & $\phi _{x^{\ast }},\phi _{u},\phi _{u_{x}}$ \\ \hline
4a & $\left\{ 
\begin{array}{c}
\phi _{u_{x}}\phi _{u}=\phi _{z-x}; \\ 
\left( \phi _{u_{x}}\right) _{k}^{\prime }\phi _{u}=\varepsilon
_{k},k=1,...,d. \\ 
\phi _{x^{\ast }}\phi _{u_{x}}=\phi _{x}.%
\end{array}%
\right. $ & --"-- \\ \hline
5 & $\left\{ 
\begin{array}{c}
\phi _{x^{\ast }}\phi _{u}=\phi _{z}; \\ 
Ft\left( gf_{x^{\ast }}\right) \phi _{u}=\varepsilon \\ 
\left( Ft\left( gf_{x^{\ast }}\right) \right) _{k}^{\prime }\phi
_{u}=\varepsilon _{k},k=1,...,d.%
\end{array}%
\right. $ & $\phi _{x^{\ast }},\phi _{u},g$ \\ \hline
6 & $\left\{ 
\begin{array}{c}
\phi _{x}=\phi _{-u}\phi _{z}; \\ 
Ft(g)\phi _{-u}=\varepsilon .%
\end{array}%
\right. $ & $\phi _{u},g$ \\ \hline
7 & $\left\{ 
\begin{array}{c}
Ft(g)\phi _{u}=\varepsilon ; \\ 
\left( Ft\left( g\right) \right) _{k}^{\prime }\phi _{u}=\varepsilon
_{k},k=1,...,d.%
\end{array}%
\right. $ & $\phi _{u},g$ \\ \hline
\end{tabular}

Notes. Notation $\left( \cdot \right) _{k}^{\prime }$ denotes the k-th
partial derivative of the function. The functions $\varepsilon $ are Fourier
transforms of the corresponding $w,$ and $\varepsilon _{k}=-iFt(w_{k})$
defined for the models in Tables 1 and 2.

Assumption 2 (that is fulfilled e.g. by generalized functions classes of
Table 3) ensures existence of solutions to the convolution equations for
models 1-7; this does not exclude multiple solutions and the next section
provides a discussion of solutions for equations in Table 4.

\subsection{Classes of solutions; support and multiplicity of solutions}

Typically, support assumptions are required to restrict multiplicity of
solutions; here we examine the dependence of solutions on the support of the
functions. The results here also give conditions under which some zeros,
e.g. in the characteristic functions, are allowed. Thus in common with e.g.
Carrasco and Florens (2010), Evdokimov and White (2011), distributions such
as the uniform or triangular for which the characteristic function has
isolated zeros are not excluded. The difference here is the extension of the
consideration of the solutions to $S^{\ast }$ and to models such as the
regression model where this approach to relaxing support assumptions was not
previously considered.

\ Recall that for a continuous function $\psi (x)$ on $R^{d}$ support is
defined as the set $W=$supp($\psi ),$ such that 
\begin{equation*}
\psi (x)=\left\{ 
\begin{array}{cc}
a\neq 0 & \text{for }x\in W \\ 
0 & \text{for }x\in R^{d}\backslash W.%
\end{array}%
\right.
\end{equation*}%
Support of a continuous function is an open set.

Generalized functions are functionals on the space $S$ and support of a
generalized function $b\in S^{\ast }$ is defined as follows (Schwartz, 1967,
p. 28). Denote by $\left( b,\psi \right) $ the value of the functional $b$
for $\psi \in S.$ Define a null set for $b\in S^{\ast }$ as the union of
supports of all functions in $S$ for which the value of the functional is
zero$:$ $\Omega =\{\cup $supp$\left( \psi \right) ,$ $\psi \in S,$ such that 
$\left( b,\psi \right) =0\}.$ Then supp$\left( b\right) =R^{d}\backslash
\Omega .$ Note that a generalized function has support in a closed set, for
example, support of the $\delta -function$ is just one point 0.

Note that for model 2 Table 4 gives the solution for $\phi _{x^{\ast }}$
directly and the inverse Fourier transform can provide the (generalized)
density function, $f_{x^{\ast }}.$

In Zinde-Walsh (2012) identification conditions in $S^{\ast }$ were given
for models 1 and 7 under assumptions that include the ones in Table 3 but
could also be more flexible.

The equations in Table 3 for models 1,3, 4, 4a, 5, 6 and 7 are of two types,
similar to those solved in Zinde-Walsh (2012). One is a convolution with one
unknown function; the other is a system of equations with two unknown
functions, each leading to the corresponding equations for their Fourier
transforms.

\subsubsection{Solutions to the equation $\protect\alpha \protect\beta =%
\protect\gamma .$}

Consider the equation 
\begin{equation}
\alpha \beta =\gamma ,  \label{product}
\end{equation}%
with one unknown function $\alpha ;$ $\beta $ is a given continuous
function. By assumption 2 the non-parametric class for $\alpha $ is such
that the equation holds in $S^{\ast }$ on $R^{d}$; it is also possible to
consider a nonparametric class for $\alpha $ with restricted support, $\bar{W%
}.$ Of course without any restrictions $\bar{W}=R^{d}.$ Recall the
differentiation operator, $\partial ^{m},$ for $m=(m_{1},...m_{d}\dot{)}$
and denote by $supp(\beta ,\partial )$ the set $\cup _{\Sigma
m_{i}=0}^{\infty }supp(\partial ^{m}\beta );$ where $supp(\partial ^{m}\beta
)$ is an open set where a continuous derivative $\partial ^{m}\beta $
exists. Any point where $\beta $ is zero belongs to this set if some
finite-order partial continuous derivative of $\beta $ is not zero at that
point (and in some open neighborhood); for $\beta $ itself $supp(\beta
)\equiv supp(\beta ,0).$

Define the functions 
\begin{equation}
\alpha _{1}=\beta ^{-1}\gamma I\left( supp(\beta ,\partial )\right) ;\alpha
_{2}(x)=\left\{ 
\begin{array}{cc}
1 & \text{for }x\in supp(\beta ,\partial ); \\ 
\tilde{\alpha} & \text{for }x\in \bar{W}\backslash (supp(\beta ,\partial ))%
\end{array}%
\right.  \label{division}
\end{equation}
with any $\tilde{\alpha}$ such that $\alpha _{1}\alpha _{2}\in Ft\left(
A\right) .$

Consider the case when $\alpha ,\beta $ and thus $\gamma $ are continuous.
For any point $x_{0}$ if $\beta (x_{0})\neq 0,$ there is a neighborhood $%
N(x_{0})$ where $\beta \neq 0,$ and division by $\beta $ is possible. If $%
\beta (x_{0})$ has a zero, it could only be of finite order and in some
neighborhood, $N(x_{0})\in supp(\partial ^{m}\beta )$ a representation 
\begin{equation}
\beta =\eta (x)\Pi _{i=1}^{d}\left( x_{i}-x_{0i}\right) ^{m_{i}}
\label{finitezero}
\end{equation}%
holds for some continuous function $\eta $ in $S^{\ast },$ such that $\eta
>c_{\eta }>0$ on $supp(\eta ).$Then $\eta ^{-1}\gamma $ in $N(x_{0})$ is a
non-zero continuous function; division of such a function by $\Pi
_{i=1}^{d}\left( x_{i}-x_{0i}\right) ^{m_{i}}$ in $S^{\ast }$ is defined
(Schwartz, 1967, pp. 125-126), thus division by $\beta $ is defined in this
neighborhood $N(x_{0})$. For the set $supp(\beta ,\partial )$ consider a
covering of every point by such neighborhoods, the possibility of division
in each neighborhood leads to possibility of division globally on the whole $%
supp(\beta ,\partial ).$ Then $a_{1}$ as defined in $\left( \ref{division}%
\right) $ exists in $S^{\ast }.$

In the case where $\gamma $ is an arbitrary generalized function, if $\beta $
is infinitely differentiable then then by (Schwartz, 1967, pp.126-127)
division by $\beta $ is defined on $supp(\beta ,\partial )$ and the solution
is given by $\left( \ref{division}\right) .$

For the cases where $\gamma $ is not continuous and $\beta $ is not
infinitely differentiable the solution is provided by%
\begin{equation*}
\alpha _{1}=\beta ^{-1}\gamma I\left( supp(\beta ,0)\right) ;\alpha
_{2}(x)=\left\{ 
\begin{array}{cc}
1 & \text{for }x\in supp(\beta ,0); \\ 
\tilde{\alpha} & \text{for }x\in \bar{W}\backslash (supp(\beta ,0))%
\end{array}%
\right.
\end{equation*}%
with any $\tilde{\alpha}$ such that $\alpha _{1}\alpha _{2}\in Ft\left(
A\right) .$

Theorem 2 in Zinde-Walsh (2012) implies that the solution to $\left( \ref%
{product}\right) $ is $a=Ft^{-1}(\alpha _{1}\alpha _{2});$ the sufficient
condition for the solution to be unique is $supp(\beta ,0)\supset \bar{W};$
if additionally either $\gamma $ is a continuous function or $\beta $ is an
infinitely continuously differentiable function it is sufficient for
uniqueness that $supp(\beta ,\partial )\supset \bar{W}.$

This provides solutions for models 1 and 6 where only equations of this type
appear.

\subsubsection{Solutions to the system of equations}

For models 3,4,5 and 7 a system of equations of the form 
\begin{eqnarray}
&&%
\begin{array}{cc}
\alpha \beta & =\gamma ; \\ 
\alpha \beta _{k}^{\prime } & =\gamma _{k},%
\end{array}
\label{twoeq} \\
k &=&1,...,d.  \notag
\end{eqnarray}%
(with $\beta $ continuously differentiable) arises. Theorem 3 in Zinde-Walsh
(2012) provides the solution and uniqueness conditions for this system of
equations. It is first established that a set of continuous functions $%
\varkappa _{k},k=1,...,d,$ that solves the equation 
\begin{equation}
\varkappa _{k}\gamma -\gamma _{k}=0  \label{difeq}
\end{equation}%
in the space $S^{\ast }$ exists and is unique on $W=supp(\gamma )$ as long
as $supp(\beta )\supset W.$ Then $\beta _{k}^{\prime }\beta ^{-1}=\varkappa
_{k}$ and substitution into $\left( \ref{difeq}\right) $ leads to a system
of first-order differential equations in $\beta .$

Case 1. Continuous functions; $W$ is an open set.

For the models 3 and 4 the system $\left( \ref{twoeq}\right) $ involves
continuous characteristic functions thus there $W$ is an open set. In some
cases $W$ can be an open set under conditions of models 5 and 7, e.g. if the
regression function is integrable in model 7.

For this case represent the open set $W$ as a union of (maximal) connected
components $\cup _{v}W_{v}.$

Then by the same arguments as in the proof of Theorem 3 in Zinde-Walsh
(2012)\ the solution can be given uniquely on $W$ as long as at some point $%
\zeta _{0v}\in (W_{v}\cap W)$ the value $\beta \left( \zeta _{0\nu }\right) $
is known for each of the connected components . Consider then $\beta
_{1}(\zeta )=\Sigma _{\nu }[\beta \left( \zeta _{0\nu }\right) \exp
\int_{\zeta _{0}}^{\zeta }\tsum\limits_{k=1}^{d}\varkappa _{k}(\xi )d\xi
]I(W_{\nu }),$ where integration is along any arc within the component that
connects $\zeta $ to $\zeta _{0\nu }.$ Then $\alpha _{1}=\beta
_{1}^{-1}\gamma ,$ and $\alpha _{2},\beta _{2}$ are defined as above by
being $1$ on $\cup _{v}W_{v}$ and arbitrary outside of this set.

When $\beta (0)=1$ as is the case for the characteristic function, the
function is uniquely determined on the connected component that includes 0.

Evdokimov and White (2012) provide a construction that permits in the
univariate case to extend the solution $\beta \left( \zeta _{0\nu }\right)
[\exp \int_{\zeta _{0}}^{\zeta }\tsum\limits_{k=1}^{d}\varkappa _{k}(\xi
)d\xi ]I(W_{\nu })$ from a connected component of support where $\beta
\left( \zeta _{0\nu }\right) $ is known (e.g. at 0 for a characteristic
function) to a contiguous connected component when on the border between the
two where $\beta =0,$ at least some finite order derivative of $\beta $ is
not zero. In the multivariate case this approach can be extended to the same
construction along a one-dimensional arc from one connected component to the
other. Thus identification is possible on a connected component of $%
supp(\beta ,\partial ).$

Case 2. $W$ is a closed set.

Generally for models 5 and 7, $W$ is the support of a generalized function
and is a closed set. It may intersect with several connected components of
support of $\beta .$ Denote by $W_{v\text{ }}$ here the intersection of a
connected component of support of $\beta $ and $W.$ Then similarly $\beta
_{1}(\zeta )=\tsum\limits_{\nu }[\beta \left( \zeta _{0\nu }\right) \exp
\int_{\zeta _{0}}^{\zeta }\tsum\limits_{k=1}^{d}\varkappa _{k}(\xi )d\xi
]I(W_{\nu }),$ where integration is along any arc within the component that
connects $\zeta $ to $\zeta _{0\nu }.$ Then $\alpha _{1}=\beta
_{1}^{-1}\varepsilon ,$ and $\alpha _{2},\beta _{2}$ are defined as above by
being $1$ on $\cup _{v}W_{v}$ and arbitrary outside of this set. The issue
of the value of $\beta $ at some point within each connected component
arises. In the case of $\beta $ being a characteristic function if there is
only one connected component, $W$ and $0\in W$ the solution is unique, since
then $\beta (0)=1.$

Note that for model 5 the solution to equations of the type $\left( \ref%
{twoeq}\right) $ would only provide $Ft(gf_{x^{\ast }})$ and $\phi _{u};$
then from the first equation for this model in Table 4 $\phi _{x^{\ast }}$
can be obtained; it is unique if supp$\phi _{x^{\ast }}=$supp$\phi _{z}$. To
solve for $g$ find $g=Ft^{-1}\left( Ft\left( gf_{x^{\ast }}\right) \right)
\cdot \left( f_{x^{\ast }}\right) ^{-1}.$

\section{Identification, partial identification and well-posedness}

\subsection{Identified solutions for the models 1-7}

As follows from the discussion of the solutions uniqueness in models
1,2,3,4,4a,5,6 holds (in a few cases up to a value of a function at a point)
if all the Fourier transforms are supported over the whole $R^{d};$ in many
cases it is sufficient that $supp(\beta ,\partial )=R^{d}.$

The classes of functions could be defined with Fourier transforms supported
on some known subset $\bar{W}$ of $R^{d},$ rather than on the whole space;
if all the functions considered have $\bar{W}$ as their support, and the
support consists of one connected component that includes 0 as an interior
point then identification for the solutions holds. For the next table assume
that $\bar{W}$ is a single connected component with $0$ as an interior
point; again $\bar{W}$ could coincide with $supp(\beta ,\partial )$. For
model 5 under Assumption B assume additionally that the value at zero: $%
Ft(gf_{x^{\ast }})(0)$ is known; similarly for model 7 under assumption B
additionally assume that $Ft(g)(0)$ is known.

Table 5. The solutions for identified models on $\bar{W}.$

\begin{tabular}{|c|c|}
\hline
Model & $%
\begin{array}{c}
\text{Solution to } \\ 
\text{equations}%
\end{array}%
$ \\ \hline
\multicolumn{1}{|l|}{$\ \ \ $1.} & \multicolumn{1}{|l|}{$\ \ \ \ \ \ \ \ \ \
\ \ \ \ \ \ \ \ \ \ \ \ \ \ \ \ \ \ f_{x^{\ast }}=Ft^{-1}\left( \phi
_{u}^{-1}\phi _{z}\right) .$} \\ \hline
2. & $f_{x^{\ast }}=Ft^{-1}\left( \phi _{-u}\phi _{z}\right) .$ \\ \hline
\multicolumn{1}{|l|}{$\ $\ 3.} & \multicolumn{1}{|l|}{$%
\begin{array}{c}
\text{Under Assumption A} \\ 
f_{x^{\ast }}=Ft^{-1}(\exp \int_{\zeta _{0}}^{\zeta
}\tsum\limits_{k=1}^{d}\varkappa _{k}(\xi )d\xi ), \\ 
\text{where }\varkappa _{k}\text{ solves }\varkappa _{k}\phi _{z}-[\left(
\phi _{z}\right) _{k}^{\prime }-\varepsilon _{k}]=0; \\ 
f_{u}=Ft^{-1}(\phi _{x^{\ast }}^{-1}\varepsilon ). \\ 
\text{Under Assumption B} \\ 
f_{u}=Ft^{-1}(\exp \int_{\zeta _{0}}^{\zeta }\tsum\limits_{k=1}^{d}\varkappa
_{k}(\xi )d\xi ); \\ 
\varkappa _{k}\text{ solves }\varkappa _{k}\phi _{z}-\varepsilon _{k}=0; \\ 
f_{x^{\ast }}=Ft^{-1}(\phi _{u}^{-1}\varepsilon ).%
\end{array}%
$} \\ \hline
4 & $%
\begin{array}{c}
f_{x^{\ast }},f_{u}\text{ obtained similarly to those in 3.;} \\ 
\phi _{u_{x}}=\phi _{x^{\ast }}^{-1}\phi _{x}.%
\end{array}%
$ \\ \hline
4a. & $%
\begin{array}{c}
f_{u_{x}},f_{u}\text{ obtained similarly to }\phi _{x^{\ast }},\phi _{u}%
\text{ in 3.;} \\ 
\phi _{x^{\ast }}=\phi _{u_{x}}^{-1}\phi _{x}.%
\end{array}%
$ \\ \hline
5. & $%
\begin{array}{c}
\text{Three steps:} \\ 
\text{1. (a) Get }Ft(gf_{x^{\ast }}),\phi _{u}\text{ similarly to }\phi
_{x^{\ast }},\phi _{u}\text{ in model 3} \\ 
\text{(under Assumption A use }Ft(gf_{x^{\ast }})(0))\text{;} \\ 
\text{2. Obtain }\phi _{x^{\ast }}=\phi _{u}^{-1}\phi _{z}; \\ 
\text{3. Get }g=\left[ Ft^{-1}\left( \phi _{x^{\ast }}\right) \right]
^{-1}Ft^{-1}(Ft(gf_{x^{\ast }})).%
\end{array}%
$ \\ \hline
6. & $\phi _{-u}=\phi _{z}^{-1}\phi _{x}$ and $g=Ft^{-1}(\phi _{x}^{-1}\phi
_{z}\varepsilon ).$ \\ \hline
7. & $%
\begin{array}{c}
\phi _{x^{\ast }},Ft(g)\text{obtained similarly to }\phi _{x^{\ast }},\phi
_{u}\text{in }3 \\ 
\text{(under Assumption A use }Ft(g)(0)).%
\end{array}%
$ \\ \hline
\end{tabular}

\subsection{Implications of partial identification.}

Consider the case of Model 1. Essentially lack of identification, say in the
case when the error distribution has characteristic function supported on a
convex domain $W_{u}$ around zero results in the solution for $\phi
_{x^{\ast }}=\phi _{1}\phi _{2},$ with $\phi _{1}$ non-zero and unique on $%
W_{u},$ and thus captures the lower-frequency components of $x^{\ast },$ and
with $\phi _{2}$ is a characteristic function of a distribution with
arbitrary high frequency components. Transforming back to densities provides
a corresponding model with independent components 
\begin{equation*}
z=x_{1}^{\ast }+x_{2}^{\ast }+u,
\end{equation*}%
where $x_{1}^{\ast }$ uniquely extracts the lower frequency part of observed 
$z.$ The more important the contribution of $x_{1}^{\ast }$ to $x^{\ast }$
the less important is lack of identification.

If the feature of interest as discussed e.g. by Matzkin (2007) involves only
low frequency components of $x^{\ast },$ it may still be fully identified
even when the distribution for $x^{\ast }$ itself is not. An example of that
is a deconvolution applied to an image of a car captured by a traffic
camera; although even after deconvolution the image may still appear blurry
the licence plate number may be clearly visible. In nonparametric regression
the polynomial growth of the regression or the expectation of the response
function may be identifiable even if the regression function is not fully
identified.

Features that are identified include any functional, $\Phi ,$ linear or
non-linear on a class of functions of interest, such that in the frequency
domain $\Phi $ is supported on $W_{u}.$

\subsection{Well-posedness in $S^{\ast }$}

Conditions for well-posedness in $S^{\ast }$ for solutions of the equations
entering in models 1-7 were established in Zinde-Walsh (2012).
Well-posedness is needed to ensure that if a sequence of functions converges
(in the topology of $S^{\ast })$ to the known functions of the equations
characterizing the models 1-7 in tables 1 and 2, then the corresponding
sequence of solutions will converge to the solution for the limit functions.
A feature of well-posedness in $S^{\ast }$ is that the solutions are
considered in a class of functions that is a bounded set in $S^{\ast }.$

The properties that differentiation is a continuous operation, and that the
Fourier transform is an isomorphism of the topological space $S^{\ast },$
make conditions for convergence in this space much weaker than those in
functions spaces, say, $L_{1},$ $L_{2}.$ Thus for density that is given by
the generalized derivative of the distribution function well-posedness holds
in spaces of generalized functions by the continuity of the differentiation
operator$.$

For the problems here however, well-posedness does not always obtain. The
main sufficient condition is that the inverse of the characteristic function
of the measurement error satisfy the condition $\left( \ref{condition}%
\right) $ with $b=\phi _{u}^{-1}$ on the corresponding support. This holds
if either the support is bounded or if the distribution is not super-smooth.
If $\phi _{u}$ has some zeros but satisfies the identification conditions so
that it has local representation $\left( \ref{finitezero}\right) $ where $%
\left( \ref{condition}\right) $ is satisfied for $b=\eta ^{-1}$
well-posedness will hold.

Example in Zinde-Walsh (2012) demonstrates that well-posedness of
deconvolution will not hold even in the weak topology of $S^{\ast }$ for
super-smooth (e.g. Gaussian) distributions on unbounded support. On the
other hand, well-posedness of deconvolution in $S^{\ast }$ obtains for
ordinary smooth distributions and thus under less restrictive conditions
than in function spaces, such as $L_{1}$ or $L_{2}$ usually considered.

In the models 3-7 with several unknown functions, more conditions are
required to ensure that all the operations by which the solutions are
obtained are continuous in the topology of $S^{\ast }.$ It may not be
sufficient to assume $\left( \ref{condition}\right) $ for the inverses of
unknown functions where the solution requires division; for continuity of
the solution the condition may need to apply uniformly.

Define a class of ordinary functions on $R^{d},$ $\Phi (m,V)$ (with $m$ a
vector of integers, $V$ a positive constant) where $b\in \Phi (m,V)$ if 
\begin{equation}
\int \Pi \left( (1+t_{i}^{2})^{-1}\right) ^{m_{i}}\left\vert b(t)\right\vert
dt<V<\infty .\text{ }  \label{condb}
\end{equation}

Then in Zinde-Walsh (2012) well-posedness is proved for model 7 as long as
in addition to Assumption A or B, for some $\Phi (m,V)$ both $\phi _{u}$ and 
$\phi _{u}^{-1}$ belong to the class $\Phi (m,V)$. This condition is
fulfilled by non-supersmooth $\phi _{u};$ this could be an ordinary smooth
distribution or a mixture with some mass point.

A convenient way of imposing well-posedness is to restrict the support of
functions considered to a bounded $\bar{W}.$ If the features of interest are
associated with low-frequency components only, then if the functions are
restricted to a bounded space the low-frequency part can be identified and
is well-posed.

\section{Implications for estimation}

\subsection{Plug-in non-parametric estimation}

Solutions in Table 5 for the equations that express the unknown functions
via known functions of observables give scope for plug-in estimation. As
seen e.g. in the example of Model 4, 4 and 4a are different expressions that
will provide different plug-in estimators for the same functions.

The functions of the observables here are characteristic functions and
Fourier transforms of density-weighted conditional expectations and in some
cases their derivatives, that can be estimated by non-parametric methods.
There are some direct estimators, e.g. for characteristic functions. In the
space $S^{\ast }$ the Fourier transform and inverse Fourier transform are
continuous operations thus using standard estimators of density weighted
expectations and applying the Fourier transform would provide consistency in 
$S^{\ast }$; the details are provided in Zinde-Walsh (2012). Then the
solutions can be expressed via those estimators by the operations from Table
5 and, as long as the problem is well-posed, the estimators will be
consistent and the convergence will obtain at the appropriate rate. As in An
and Hu (2012), the convergence rate may be even faster for well-posed
problems in $S^{\ast }$ than the usual nonparametric rate in (ordinary)
function spaces. For example, as demonstrated in Zinde-Walsh (2008) kernel
estimators of density that may diverge if the distribution function is not
absolutely continuous, are always (under the usual assumptions on
kernel/bandwidth) consistent in the weak topology of the space of
generalized functions, where the density problem is well-posed. Here,
well-posedness holds for deconvolution as long as the error density is not
super-smooth.

\subsection{Regularization in plug-in estimation}

When well-posedness cannot be ensured, plug-in estimation will not provide
consistent results and some regularization is required; usually spectral
cut-off is employed for the problems considered here. In the context of
these non-parametric models regularization requires extra information: the
knowledge of the rate of decay of the Fourier transform of some of the
functions.

For model 1 this is not a problem since $\phi _{u}$ is assumed known; the
regularization uses the information about the decay of this characteristic
function to construct a sequence of compactly supported solutions with
support increasing at a corresponding rate. In $S^{\ast }$ no regularization
is required for plug-in estimation unless the error distribution is
super-smooth. Exponential growth in $\phi _{u}^{-1}$ provides a logarithmic
rate of convergence in function classes for the estimator (Fan, 1991). Below
we examine spectral cut-off regularization for the deconvolution in $S^{\ast
}$ when the error density is super-smooth.

With super-smooth error in $S^{\ast }$ define a class of generalized
functions $\Phi (\Lambda ,m,V)$ for some non-negative-valued function $%
\Lambda $; a generalized function $b\in \Phi (\Lambda ,m,V)$ if there exists
a function $\bar{b}(\zeta )\in \Phi (m,V)$ such that also $\bar{b}(\zeta
)^{-1}\in \Phi (m,V)$ and $b=\bar{b}(\zeta )\exp \left( -\Lambda (\zeta
)\right) .$ Note that a linear combination of functions in $\Phi (\Lambda
,m,V)$ belongs to the same class. Define convergence: a sequence of $%
b_{n}\in \Phi (\Lambda ,m,V)$ converges to zero if the corresponding
sequence $\bar{b}_{n}$ converges to zero in $S^{\ast }.$

Convergence in probability for a sequence of random functions, $\varepsilon
_{n},$ in $S^{\ast }$ is defined as follows: $(\varepsilon _{n}-\varepsilon
)\rightarrow _{p}0$ in $S^{\ast }$ if for any set $\psi _{1},...,\psi
_{v}\in S$ the random vector of the values of the functionals converges: $%
\left( (\varepsilon _{n}-\varepsilon ,\psi _{1}),...,(\varepsilon
_{n}-\varepsilon ,\psi _{v})\right) \rightarrow _{p}0.$

\textbf{Lemma 2.} \textit{If in model 1 }$\phi _{u}=b\in \Phi (\Lambda ,m,V),
$\textit{\ where }$\Lambda $\textit{\ is a polynomial function of order no
more than }$k,$\textit{\ and }$\varepsilon _{n}$\textit{\ is a sequence of
estimators of }$\varepsilon $\textit{\ that are consistent in }$S^{\ast
}:r_{n}(\varepsilon _{n}-\varepsilon )\rightarrow _{p}0$\textit{\ in }$%
S^{\ast }$\textit{\ at some rate }$r_{n}\rightarrow \infty ,$\textit{\ then
for any sequence of constants }$\bar{B}_{n}:$\textit{\ }$0<\bar{B}%
_{n}<\left( \ln r_{n}\right) ^{\frac{1}{k}}$\textit{\ and the corresponding
set }$B_{n}=\left\{ \zeta :\left\Vert \zeta \right\Vert <\bar{B}_{n}\right\} 
$\textit{\ the sequence of regularized estimators }$\phi
_{u}^{-1}(\varepsilon _{n}-\varepsilon )I(B_{n})$\textit{\ converges to zero
in probability in }$S^{\ast }.$\textit{\ }

Proof. For $n$ the value of the random functional 
\begin{equation*}
(\phi _{u}^{-1}(\varepsilon _{n}-\varepsilon )I(B_{n}),\psi )=\int \bar{b}%
^{-1}(\zeta )r_{n}(\varepsilon _{n}-\varepsilon )r_{n}^{-1}I(B_{n})\exp
\left( \Lambda (\zeta )\right) \psi (\zeta )d\zeta .
\end{equation*}%
Multiplication by $\bar{b}^{-1}\in \Phi (m,V),$ that corresponds to $\phi
_{u}=b$ does not affect convergence thus $\bar{b}^{-1}(\zeta
)r_{n}(\varepsilon _{n}-\varepsilon )$ converges to zero in probability in $%
S^{\ast }.$ To show that $(\phi _{u}^{-1}(\varepsilon _{n}-\varepsilon
)I(B_{n}),\psi )$ converges to zero it is sufficient to show that the
function $r_{n}^{-1}I(B_{n})\exp \left( \Lambda (\zeta )\right) \psi (\zeta
) $ is bounded$.$ It is then sufficient to find $B_{n}$ such that $%
r_{n}^{-1}I(B_{n})\exp \left( \Lambda (\zeta )\right) $ is bounded (by
possibly a polynomial), thus it is sufficient that $\underset{B_{n}}{\sup }%
\left\vert \exp \left( \Lambda (\zeta )\right) r_{n}^{-1}\right\vert $ be
bounded. This will hold if $\exp \left( \bar{B}_{n}^{k}\right) <r_{n},$ $%
\bar{B}_{n}^{k}<\ln r_{n}.\blacksquare $

Of course an even slower growth for spectral cut-off would result from $%
\Lambda $ that grows faster than a polynomial. The consequence of the slow
growth of the support is usually a correspondingly slow rate of convergence
for $\phi _{u}^{-1}\varepsilon _{n}I(B_{n}).$ Additional conditions (as in
function spaces) are needed for the regularized estimators to converge to
the true $\gamma $.

It may be advantageous to focus on lower frequency components and ignore the
contribution from high frequencies when the features of interest depend on
the contribution at low frequency.

\section{Concluding remarks}

Working in spaces of generalized functions extends the results on
nonparametric identification and well-posedness for a wide class of models.
Here identification in deconvolution is extended to generalized densities in
the class of all distributions from the usually considered classes of
integrable density functions. In regression with Berkson error nonparametric
identification in $S^{\ast }$\ holds for functions of polynomial growth,
extending the usual results obtained in $L_{1};$ a similar extension applies
to regression with measurement error and Berkson type measurement; this
allows to consider binary choice and polynomial regression models. Also,
identification in models with sum-of-peaks regression function that cannot
be represented in function spaces is included. Well-posedness results in $%
S^{\ast }$ also extend the results in the literature provided in function
spaces; well-posedness of deconvolution holds as long as the characteristic
function of the error distribution does not go to zero at infinity too fast
(as e.g. super-smooth) and a similar condition provides well-posedness in
the other models considered here.

Further investigation of the properties of estimators in spaces of
generalized functions requires deriving the generalized limit process for
the function being estimated and investigating when it can be described as a
generalized Gaussian process. A generalized Gaussian limit process holds for
kernel estimator of the generalized density function (Zinde-Walsh, 2008).
Determining the properties of inference based on the limit process for
generalized random functions requires both further theoretical development
and simulations evidence.


\begin{thebibliography}{99}
\bibitem{} An, Y, and Y. Hu (2012) Well-posedness of measurement error
models for self-reported data, Journal of Econometrics, vol. 168 (2012), pp.
259-269.

\bibitem{} Ben-Moshe, D. (2012) Identification of Dependent Multidimensional
Unobserved Variables in a System of Linear Equations, working paper, UCLA.

\bibitem{} Bonhomme, S. and J-M. Robin (2010) Generalized nonparametric
deconvolution with an application to earnings dynamics, Review of Economic
Studies, 77, pp. 491-533.

\bibitem{} Butucea, C. and M.L.Taupin (2008) New M-estimators in
semi-parametric regression with errors in variables, Annales de l'Institut
Henri Poincar\'{e} - Probabilit\'{e}s et Statistiques,V. 44, pp. 393--421.

\bibitem{cunha} Carroll, R. J., Ruppert, D., Stefanski, L. A. and C.M.
Crainiceanu (2006). Measurement Error in Nonlinear Models: A Modern
Perspective. Chapman \& Hall.

\bibitem{} Carrasco, M. and J.-P. Florens (2010), A Spectral Method for
Deconvolving a Density, Econometric Theory, 27 , pp 546-581.\ 

\bibitem{} Chen, X., H. Hong and D.Nekipelov (2011), Nonlinear models of
measurement errors, Journal of Economic Literature, 49, pp.901-937.

\bibitem{} Chen, X., Hong, H. and Tamer, E. (2005), Measurement Error Models
with Auxiliary Data, Review of Economic Studies, 72, pp. 343--366.

\bibitem{} Cunha, F., J.J. Heckman and S.M.Schennach (2010), Estimating the
Technology of Cognitive and Noncognitive Skill Formation, Econometrica,
v.78, 883-933.

\bibitem{} Evdokimov, K. (2011), Identification and Estimation of a
Nonparametric Panel Data Model with Unobserved Heterogeneity, working paper.

\bibitem{} Evdokimov, K. and H. White (2011) An Extension of a Lemma of
Kotlyarski, Econometric Theory, forthcoming.

\bibitem{} Fan, J. Q. (1991), On the Optimal Rates of Convergence for
Nonparametric Deconvolution Problems, Annals of Statistics, 19, pp.
1257-1272.

\bibitem{} Green, D.A. and \ W.C. Riddell (1997) Qualifying for unemployment
insurance: An empirical analysis, Economic Journal, 107, pp. 67-84.

\bibitem{had} Hadamard, J. (1923): Lectures on Cauchy's Problem in Linear
Partial Differential Equations. Yale University Press, New Heaven.

\bibitem{} Hu, Y. (2008), Identification and Estimation of Nonlinear Models
with Misclassification Error using Instrumental Variables: A General
Solution, Journal of Econometrics, 144, pp. 27-61.

\bibitem{} Hu, Y. and G. Ridder (2010), On Deconvolution as a First Stage
Nonparametric Estimator, Econometric Reviews, 29, pp. 1-32.

\bibitem{} Karlin, S. (1959) The Theory of Infinite Games, v.II,
Addison-Wesley Publ.

\bibitem{} Kotlyarski, I. (1967), On Characterizing the Gamma and Normal
Distribution, Pacific Journal of Mathematics, 20, pp. 69-76.

\bibitem{} Li., T. (2002) Robust and consistent estimation of nonlinear
errors-in-variables models, Journal of Econometrics, 110, pp.1-26.

\bibitem{L} Li, T. and Ch. Hsiao (2004) Robust estimation of generalized
models with measurement error, Journal of Econometrics, 118, pp. 51-65.

\bibitem{} Li, T. and Vuong, Q. (1998), Nonparametric Estimation of the
Measurement Error Model Using Multiple Indicators, Journal of Multivariate
Analysis, 65, 139-165.

\bibitem{} Mahajan, A. (2006) Identification and estimation of regression
models with misclassification, Econometrica, 74, pp. 631-665.

\bibitem{} Matzkin, R.L. (2007) Nonparametric Identification,\ Chapter 73 in
Handbook of Econometrics, Vol. 6b, edited by J.J. Heckman and E.E. Leamer,
Elsevier B.V., 5307-5368.

\bibitem{} A.Meister (2009) Deconvolution problems in nonparametric
statistics, Lecture notes in statistics, Springer-Verlag.

\bibitem{Newey} Newey, W. (2001), Flexible Simulated Moment Estimation of
Nonlinear Errors-in-Variables Models, Review of Economics and Statistics, v.
83, 616-627.

\bibitem{} Schennach, S. (2004) Nonparametric regression in the presence of
measurement error, Econometric Theory, 20, pp. 1046-1093.

\bibitem{} Schennach, S. (2007) Instrumental variable estimation in
nonlinear errors-in-variables models, Econometrica, v.75, pp. 201-239.

\bibitem{sz} Schwartz, L. (1966) "Th\'{e}orie des distributions", Hermann,
Paris.

\bibitem{} Taupin, M.-L. (2001) Semi-parametric estimation in the nonlinear
structural errors-in-variables model. Ann. Statist., 29, pp. 66--93.

\bibitem{Wa} Wang, L. Estimation of nonlinear models with Berkson
measurement errors, Annals of Statistics, 32, pp. 2559-2579.

\bibitem{} Zinde-Walsh, V. (2008) Kernel estimation when density may not
exist, Econometric Theory, v.24, 696-725.

\bibitem{zweiv} Zinde-Walsh, V. (2009), Errors-in-variables models: a
generalized functions approach, working paper, arXiv:0909.5390v1 [stat.ME],
McGill University working paper.

\bibitem{} Zinde-Walsh, V. (2011), Presidential Address: Mathematics in
economics and econometrics, Canadian Journal of Economics, v.44, pp.
1052-1068.

\bibitem{} Zinde-Walsh, V. (2012), Measurement error and deconvolution in
spaces of generalized functions, arXiv:1009.4217v2 [math.ST]; earlier
version (2010): arXiv:1009.4217v1[MATH.ST].
\end{thebibliography}
\end{document}